\documentclass[%
 reprint,
 amsmath,amssymb,
 aip,
]{revtex4-1}

\usepackage{graphicx}
\usepackage{dcolumn}
\usepackage{bm}
\usepackage{mhchem}
\usepackage{color}

\begin{document}


\title{Development of a Practical Multicomponent Density Functional for Electron-Proton Correlation to Produce Accurate Proton Densities}

\author{Yang Yang}
 \affiliation{Department of Chemistry, University of Illinois at Urbana-Champaign, 600 South Mathews Ave, Urbana, Illinois 61801, U.S.}
 \author{Kurt R. Brorsen}
 \affiliation{Department of Chemistry, University of Illinois at Urbana-Champaign, 600 South Mathews Ave, Urbana, Illinois 61801, U.S.}
 \author{Tanner Culpitt}
 \affiliation{Department of Chemistry, University of Illinois at Urbana-Champaign, 600 South Mathews Ave, Urbana, Illinois 61801, U.S.}
 \author{Michael V. Pak}
 \affiliation{Department of Chemistry, University of Illinois at Urbana-Champaign, 600 South Mathews Ave, Urbana, Illinois 61801, U.S.}
\author{Sharon Hammes-Schiffer}%
 \email{shs3@illinois.edu}
\affiliation{Department of Chemistry, University of Illinois at Urbana-Champaign, 600 South Mathews Ave, Urbana, Illinois 61801, U.S.}%

\date{\today}

\begin{abstract}
Multicomponent density functional theory (DFT) enables the consistent quantum mechanical treatment of both electrons and protons.  A major challenge has been the design of electron-proton correlation functionals that produce even qualitatively accurate proton densities.  Herein an electron-proton correlation functional, epc17, is derived analogously to the Colle-Salvetti formalism for electron correlation and is implemented within the nuclear-electronic orbital (NEO) framework.  The NEO-DFT/epc17 method produces accurate proton densities efficiently and is promising for diverse applications.
\end{abstract}

\maketitle


\section{Introduction}

Density functional theory (DFT) \cite{Hohenberg_1964_864, Parr_1994_} is a powerful computational tool in chemistry and physics. Typically DFT calculations are conducted in the context of the Born-Oppenheimer approximation, where the nuclei are treated as fixed point charges when solving the electronic Schr{\"o}dinger equation. However, the quantum mechanical nature of the nuclei and the non-Born-Oppenheimer effects between electrons and nuclei are important for many challenging physical and chemical problems, such as proton-coupled electron transfer \cite{Hammes-Schiffer_2015_8860}. Multicomponent DFT, which enables the quantum mechanical treatment of more than one type of particle, such as electrons and protons, was developed to tackle these problems \cite{Capitani_1982_568, Shigeta_1998_659, Kreibich_2001_2984, Chakraborty_2008_153001}. Analogs to the Hohenberg-Kohn theorems \cite{Hohenberg_1964_864} have been proven for multicomponent systems \cite{Capitani_1982_568, Kreibich_2001_2984, Kreibich_2008_22501}, and the Kohn-Sham formalism \cite{Kohn_1965_1133}, in which the kinetic energy functional is approximated with electronic and nuclear orbitals, has been extended to multicomponent systems \cite{Shigeta_1998_659, Udagawa_2006_244105, Pak_2007_4522, Kreibich_2001_2984, Kreibich_2008_22501, Chakraborty_2008_153001, Chakraborty_2009_124115}. This paper centers on the nuclear-electronic orbital (NEO) \cite{Webb_2002_4106} implementation of multicomponent DFT, denoted NEO-DFT \cite{Pak_2007_4522, Chakraborty_2008_153001}, applied to systems in which specific hydrogen nuclei are treated quantum mechanically, and at least two other nuclei are treated as classical point charges to avoid difficulties with translations and rotations.

A major challenge in the Kohn-Sham implementation of multicomponent DFT is the design of accurate and practical electron-proton correlation functionals. When electron-proton correlation is neglected, the proton densities are highly over-localized, leading to unphysical results for most properties of interest. Efforts to approximate the electron-proton correlation functional by analyzing the connectivity between the electron-proton pair density and the total wave function or certain orbitals \cite{Kreibich_2001_2984, Chakraborty_2008_153001, Sirjoosingh_2011_2689, Sirjoosingh_2012_174114, Kreibich_2008_22501} have improved proton densities but are computationally expensive or not easily transferable to general systems. Recent attempts \cite{Imamura_2008_735, Udagawa_2014_52519} to develop simple electron-proton correlation functionals related to the Colle-Salvetti formulation \cite{Colle_1975_329}, which forms the basis of the Lee-Yang-Parr (LYP) electron-electron correlation functional \cite{Lee_1988_785}, relied on problematic approximations. For example, Ref. \cite{Imamura_2008_735} has the incorrect sign of electron-proton correlation in the paper and incorrectly assumed a norm condition that leads to a singularity problem. Moreover, these previous attempts only applied those functionals as energetic corrections after the self-consistent-field (SCF) procedure. As a result, these approaches  \cite{Imamura_2008_735, Udagawa_2014_52519} do not influence the highly over-localized proton densities and thus are not useful for describing any properties that rely on the proton density. In this paper, we develop an electron-proton correlation functional based on the Colle-Salvetti formulation using well-justified approximations and apply this new functional within the SCF procedure, producing accurate proton densities at a low computational cost. 

\section{Theory}

Consider a multicomponent system with $N_e$ electrons and $N_p$ protons, where the subscripts $e$ and $p$ denote electrons and protons for all variables hereafter. Within the framework of multicomponent DFT, the total energy can be expressed as
\begin{equation}
\label{eq:DFTenergy}
\begin{aligned}
E[\rho_e,\rho_p]=&(T_s[\rho_e]+T_s[\rho_p])+(V_e^{ext}[\rho_e]+V_p^{ext}[\rho_p])\\&+(J_{ee}[\rho_e]+J_{pp}[\rho_p]+J_{ep}[\rho_e,\rho_p])\\&+(E_{e}^{xc}[\rho_e]+E_{p}^{xc}[\rho_p])+(E^{epc}[\rho_e,\rho_p]),
\end{aligned}
\end{equation}
where the quantities in each parenthesis represent the non-interacting reference kinetic energy, external potential energy, mean-field Coulomb interactions, same-particle exchange-correlation (xc) energy, and electron-proton correlation (epc) energy, respectively. The electron and proton densities are defined in terms of the electron and proton orbitals, $\phi_{e}^{i}(\mathbf{r}_e)$ and $\phi_{p}^{I}(\mathbf{r}_p)$:
	$\rho_e(\mathbf{r}_e)=\sum_i^{N_e}|\phi_e^i(\mathbf{r}_e)|^2$,
	$\rho_p(\mathbf{r}_p)=\sum_{I}^{N_p}|\phi_p^{I}(\mathbf{r}_p)|^2$.
The Kohn-Sham equations for electrons and protons obtained by minimizing the total energy are
\begin{subequations}
	\begin{align}
	\bigg(-\frac{1}{2}\nabla^2_e+v_e(\mathbf{r}_e)\bigg)\phi_e^i(\mathbf{r}_e)=&\varepsilon_e^i\phi_e^i(\mathbf{r}_e),\\
	\bigg(-\frac{1}{2m_p}\nabla^2_p+v_p(\mathbf{r}_p)\bigg)\phi_p^{I}(\mathbf{r}_p)=&\varepsilon_p^{I}\phi_p^{I}(\mathbf{r}_p),
	\end{align}
\end{subequations}
in atomic units with the effective potentials for the electrons and protons defined as
\begin{subequations}
	\begin{align}
	v_e(\mathbf{r}_e)=&v_e^{ext}(\mathbf{r}_e)+v_e^{J_{ee}}(\mathbf{r}_e)+v_e^{J_{ep}}(\mathbf{r}_e)\\&\notag+v_e^{xc}(\mathbf{r}_e)+v_e^{epc}(\mathbf{r}_e),\\
	v_p(\mathbf{r}_p)=&v_p^{ext}(\mathbf{r}_p)+v_p^{J_{pp}}(\mathbf{r}_p)+v_p^{J_{ep}}(\mathbf{r}_p)\\&\notag+v_p^{xc}(\mathbf{r}_p)+v_p^{epc}(\mathbf{r}_p).
	\end{align}
\end{subequations}
Each potential term is obtained by taking the functional derivative of the corresponding energy term in Eq. \eqref{eq:DFTenergy} with respect to either the electron or the proton density. The electron-electron exchange-correlation functional is defined the same as in standard electronic DFT, and previously developed functionals can be used. Because quantum protons are spatially localized in molecular systems, the proton-proton exchange and correlation energies are negligible, and can be approximated with the diagonal Hartree-Fock exchange terms to eliminate self-interaction error. For systems with multiple protons, the proton orbitals can be assumed to be only singly occupied with high spin.

As discussed above, the epc functional $E^{epc}[\rho_e,\rho_p]$ is the most critical term in the multicomponent Kohn-Sham formalism. The approach that completely neglects this term produces highly over-localized proton densities and therefore is not a useful tool for accurately describing the quantum behavior of protons. Herein we develop a multicomponent epc functional, denoted epc17, based on an analog of the Colle-Salvetti formulation but utilizing different approximations from previous attempts \cite{Imamura_2008_735, Udagawa_2014_52519} in an effort to produce accurate proton densities. The epc17 functional is a multicomponent counterpart to the well-known LYP functional \cite{Lee_1988_785} in conventional electronic DFT, excluding the density gradient terms in its current form, and it is implemented within the SCF procedure.

The derivation of our density functional begins with an analog of the Colle-Salvetti ansatz \cite{Colle_1975_329}, which approximates the total wave function for a multicomponent electron-proton system as:
\begin{equation}
\label{eq:CSansatz}
\begin{aligned}
&\Psi(\mathbf{x}_{1,e},\mathbf{x}_{2,e},\dots\mathbf{x}_{N_e,e},\mathbf{r}_{1,p},\mathbf{r}_{2,p},\dots\mathbf{r}_{N_p,p})\\=&\Psi^{FCI}_{e}(\mathbf{x}_{1,e},\mathbf{x}_{2,e},\dots\mathbf{x}_{N_e,e})\Psi^{FCI}_{p}(\mathbf{r}_{1,p},\mathbf{r}_{2,p},\dots\mathbf{r}_{N_p,p})\\
&\cdot\prod_{i\in N_e,\,I\in N_p}[1-\varphi(\mathbf{r}_{i,e},\mathbf{r}_{I,p})],
\end{aligned}
\end{equation}
where $\mathbf{x}_{i,e}$ denotes the combined electronic variable for the spatial coordinate $\mathbf{r}_{i,e}$ and spin $\sigma_{i,e}$, and only the spatial coordinate $\mathbf{r}_{I,p}$ is used for protons as the nuclei are assumed to be high spin. FCI denotes full configuration interaction for a particular type of particle, which takes into account all of the exchange and correlation effects between particles of this type but only includes a mean-field Coloumb potential from the other type of particle. Therefore, the product of the electron and proton FCI wave functions (i.e., Eq. \eqref{eq:CSansatz} without the last correlation factor) is the mean-field FCI electron-proton wave function \cite{Cassam-Chenaie_2017_52}. The electron-proton correlation effects are included in the last Jastrow factor, with the function $\varphi(\mathbf{r}_{i,e},\mathbf{r}_{I,p})$ defined as
\begin{equation}
\label{eq:CSfactor}
\begin{aligned}
\varphi(\mathbf{r}_{i,e},\mathbf{r}_{I,p})=e^{-\beta^{2}(\mathbf{R})r^{2}}[1-\xi(\mathbf{R})(1-r)],
\end{aligned}
\end{equation}
where $r=|\mathbf{r}|=|\mathbf{r}_{i,e}-\mathbf{r}_{I,p}|$, and $\mathbf{R}$ is the center of mass for the two particles, which is approximately the same as $\mathbf{r}_{I,p}$ because of the significantly larger mass of the proton compared to the electron. In this equation, $\beta(\mathbf{R})$ and $\xi(\mathbf{R})$ are two functions that will be determined below. This ansatz is chosen because it fulfills the electron-proton cusp condition as $\mathbf{r}_{i,e}\rightarrow\mathbf{r}_{I,p}$ \cite{Bingel_1967_54} and reduces to the exact mean-field FCI electron-proton wave function when the electrons and protons are far apart.

Analogous to the original Colle-Salvetti formulation for electron-electron correlation, the electron-proton pair density matrix, which was the second-order reduced density matrix in the electronic case, can be approximated by the product of the mean-field pair density matrix and a correlation factor,
\begin{equation}
\label{eq:PairDensity}
\begin{aligned}
P_{2,ep}(\mathbf{x}_{1,e}^{\prime},\mathbf{r}_{1,p}^{\prime};&\mathbf{x}_{1,e},\mathbf{r}_{1,p})\approx P_{2,ep}^{FCI}(\mathbf{x}_{1,e}^{\prime},\mathbf{r}_{1,p}^{\prime};\mathbf{x}_{1,e},\mathbf{r}_{1,p})\\&\cdot[1-\varphi(\mathbf{r}_{1,e}^{\prime},\mathbf{r}_{1,p}^{\prime})][1-\varphi(\mathbf{r}_{1,e},\mathbf{r}_{1,p})],
\end{aligned}
\end{equation}
where the mean-field electron-proton pair density matrix is the product of the associated electron and proton first-order reduced density matrices: $P_{2,ep}^{FCI}(\mathbf{x}_{1,e}^{\prime},\mathbf{r}_{1,p}^{\prime};\mathbf{x}_{1,e},\mathbf{r}_{1,p})=P_{1,e}^{FCI}(\mathbf{x}_{1,e}^{\prime};\mathbf{x}_{1,e})P_{1,p}^{FCI}(\mathbf{r}_{1,p}^{\prime};\mathbf{r}_{1,p})$.
The electron-proton correlation energy is formally defined as the difference between the exact total energy associated with the wave function in Eq. \eqref{eq:CSansatz} and the energies associated with the electron and proton mean-field FCI wave functions after removing the double counting of the mean-field electron-proton Coulomb interaction energy:
	\begin{equation}
	\label{eq:EcorrelationDef}
	\begin{aligned}
	E^{epc}=&E-(E^{FCI}_{e}+E^{FCI}_{p}-J_{ep})\\
	=&\mathrm{Tr[}(-\frac{1}{2}\nabla_{e}^{2}+v_{e}^{ext})(P_{1,e}-P_{1,e}^{FCI})]+\mathrm{Tr[}V_{ee}(P_{2,e}\\&-P_{2,e}^{FCI})]+\mathrm{Tr[}(-\frac{1}{2m_{p}}\nabla_{p}^{2}+v_{p}^{ext})(P_{1,p}-P_{1,p}^{FCI})]\\&+\mathrm{Tr[}V_{pp}(P_{2,p}-P_{2,p}^{FCI})]-\int\mathrm{d}\mathbf{x}_{1,e}\mathrm{d}\mathbf{r}_{1,p}\frac{1}{r}\\&\cdot[P_{2,ep}(\mathbf{x}_{1,e},\mathbf{r}_{1,p})-P_{2,ep}^{FCI}(\mathbf{x}_{1,e},\mathbf{r}_{1,p})].
	\end{aligned}
	\end{equation}
Here $V_{ee}$ and $V_{pp}$ are two-particle Coulomb repulsion operators,  $P_{1,e}$ and $P_{2,e}$ are the electron first- and second-order reduced density matrices associated with the total wave function in Eq. \eqref{eq:CSansatz}, and $P_{1,p}$ and $P_{2,p}$ are the proton first- and second-order reduced density matrices defined analogously. Moreover, the pair density $P_{2,ep}(\mathbf{x}_{1,e},\mathbf{r}_{1,p})$ denotes the diagonal elements of the electron-proton pair density matrix. 
Assuming that the electron and proton first- and second-order reduced density matrices are reasonably well approximated by those from mean-field FCI, the electron-proton correlation energy reduces to the last term in Eq. \eqref{eq:EcorrelationDef}:
\begin{equation}
\label{eq:EcorrelationSimp}
\begin{aligned}
E^{epc}
=&-\int\mathrm{d}\mathbf{x}_{1,e}\mathrm{d}\mathbf{r}_{1,p}\frac{1}{r}P_{2,ep}^{FCI}(\mathbf{x}_{1,e},\mathbf{r}_{1,p})\\&\cdot[\varphi^2(\mathbf{r}_{1,e},\mathbf{r}_{1,p})-2\varphi(\mathbf{r}_{1,e},\mathbf{r}_{1,p})].
\end{aligned}
\end{equation}
Equating the electron and proton densities associated with the full wave function in Eq. \eqref{eq:CSansatz} to those associated with the mean-field FCI electron-proton wave function, which is a direct consequence of the above approximation for the second-order reduced density matrices, leads to an approximate relationship between the undetermined functions $\xi(\mathbf{R})$ and $\beta(\mathbf{R})$:
\begin{widetext}
\begin{equation}
\label{eq:ApproxPhi}
\begin{aligned}
\xi(\mathbf{R})	\approx	\frac{-9.2 \beta^2(\mathbf{R})+14.1 \beta(\mathbf{R})\sqrt{1.0 \beta^2(\mathbf{R})-2.0 \beta(\mathbf{R})+1.2} +12.0 \beta(\mathbf{R})}{5.0 \beta^2(\mathbf{R})-8.0 \beta(\mathbf{R})+3.8}.
\end{aligned}
\end{equation}
\end{widetext}

In the original electronic Colle-Salvetti formulation, $\beta(\mathbf{R})$ is assumed to be inversely proportional to the Wigner-Seitz radius $r_{s,e}\propto\rho_e^{-1/3}(\mathbf{R})$ from a uniform electron gas model. In the multicomponent electron-proton case, however, $\beta(\mathbf{R})$ must depend on both the electron and proton densities. Thus, we define a geometric mean Wigner-Seitz radius $r_{s,ep}=(r_{s,e}r_{s,p})^{1/2}$, where $r_{s,p}$ is defined analogously to $r_{s,e}$, and assume $\beta(\mathbf{R})$ to be inversely proportional to the local mean $r_{s,ep}(\mathbf{R})$
\begin{equation}
\label{eq:CorrelationLength}
\begin{aligned}
\beta(\mathbf{R})\propto\frac{1}{r_{s,ep}(\mathbf{R})}\propto\rho_e^{1/6}(\mathbf{R})\rho_p^{1/6}(\mathbf{R}).
\end{aligned}
\end{equation}
Note that Refs. \cite{Imamura_2008_735,Udagawa_2014_52519} assumed that $\beta(\mathbf{R})\propto\rho_e^{1/3}(\mathbf{R})$ in deriving their expressions for electron-proton correlation energy; however, such neglect of the dependence on the proton density does not seem to be physically reasonable.

Although $\xi(\mathbf{R})$ and $\beta(\mathbf{R})$ have been determined, the epc functional in Eq. \eqref{eq:EcorrelationSimp} is still not practical because it involves an expensive six-dimensional integration over both electron and proton coordinates. We continue following the strategy of the Colle-Salvetti formalism and transform the coordinates \{$\mathbf{x}_{1,e},\mathbf{r}_{1,p}$\} to \{$\mathbf{R}$,$\mathbf{r}$\}. After making use of the mathematical relation
\begin{equation}
\label{eq:EqualityRelation}
\begin{aligned}
\int\mathrm{d}\mathbf{r}&\frac{e^{-\beta^2r^2}}{r}F(\mathbf{r})=4\pi F(0)\int re^{-\beta^2r^2}\mathrm{d}r\\&+\frac{2\pi}{3}(\nabla_{r}^{2}F(\mathbf{r}))|_{r=0}\int r^{3}e^{-\beta^2r^2}\mathrm{d}r+O(r^{6}),
\end{aligned}
\end{equation}
and Eq. \eqref{eq:CSfactor}, the electron-proton correlation energy becomes
\begin{widetext}
\begin{equation}
\label{eq:Ecorr}
\begin{aligned}
E^{epc}
\approx&	-4\pi\int\mathrm{d}\mathbf{R}P_{2,ep}^{FCI}(\mathbf{R},0)\int\mathrm{d}re^{-2\beta^{2}(\mathbf{R})r^{2}}[1-\xi(\mathbf{R})(1-r)]^{2}r\\
&+8\pi\int\mathrm{d}\mathbf{R}P_{2,ep}^{FCI}(\mathbf{R},0)\int\mathrm{d}re^{-\beta^{2}(\mathbf{R})r^{2}}[1-\xi(\mathbf{R})(1-r)]r\\
&-2\pi/3\int\mathrm{d}\mathbf{R}(\nabla_{r}^{2}P_{2,ep}^{FCI}(\mathbf{R},\mathbf{r}))|_{r=0}\int\mathrm{d}re^{-2\beta^{2}(\mathbf{R})r^{2}}[1-\xi(\mathbf{R})(1-r)]^{2}r^{3}\\
&+4\pi/3\int\mathrm{d}\mathbf{R}(\nabla_{r}^{2}P_{2,ep}^{FCI}(\mathbf{R},\mathbf{r}))|_{r=0}\int\mathrm{d}re^{-\beta^{2}(\mathbf{R})r^{2}}[1-\xi(\mathbf{R})(1-r)]r^{3}.
\end{aligned}
\end{equation}
\end{widetext}
In this paper, we neglect the last two terms involving the second-order derivatives of the pair density and leave the inclusion of these terms for future development. Therefore, we are able to obtain a simple form for the epc functional. Substituting Eq. \eqref{eq:ApproxPhi} into the first two terms of Eq. \eqref{eq:Ecorr} and integrating leads to the final form of our epc17 functional, which we approximate as:
	\begin{equation}
	\label{eq:FunctionalForm}
	\begin{aligned}
	E^{epc}=-\int\mathrm{d}\mathbf{R}\frac{\rho_{e}(\mathbf{R})\rho_{p}(\mathbf{R})}{a-b\rho_{e}^{1/2}(\mathbf{R})\rho_{p}^{1/2}(\mathbf{R})+c\rho_{e}(\mathbf{R})\rho_{p}(\mathbf{R})},
	\end{aligned}
	\end{equation}
with three parameters to be determined below. Although the derivation of the epc17 functional is based on the wave function in Eq. \eqref{eq:CSansatz} and its associated density matrices, these are never explicitly constructed. This simple type of functional has only two fundamental variables, namely electron density $\rho_e$ and proton density $\rho_p$. Therefore, it is a local density approximation (LDA) form for both electrons and protons. Note that $\rho_e$ is the total electron density for both closed-shell and open-shell cases, and $\rho_p$ is the total proton density, where the protons are assumed to be high spin.

\section{results}

We have implemented the epc17 functional in GAMESS \cite{Schmidt_1993_1347} and applied this approach to the molecules \ce{FHF-} and \ce{HCN}, which have been extensively studied for testing purposes \cite{Sirjoosingh_2015_214107, Brorsen_2015_214108, Culpitt_2016_44106}.  The heavy nuclei are treated as point charges fixed to the positions obtained by optimizing the geometry at the conventional electronic DFT level, and the hydrogen nucleus, as well as all electrons, are treated quantum mechanically. The def2-QZVP \cite{Weigend_2005_3297} electronic basis set and an even-tempered 8{\it s}8{\it p}8{\it d} proton basis set centered at the optimized hydrogen position with exponents running from $2\sqrt{2}$ to $32$ are used for all calculations. The B3LYP functional \cite{Becke_1988_3098, Lee_1988_785} is used for electron-electron exchange-correlation. Our epc17 functional with parameters $a$, $b$, and $c$ optimized to be 2.35, 2.4, and 3.2, is used for electron-proton correlation. Note that the parameters are fit to accurate proton densities for \ce{FHF-}, analogous to the Colle-Salvetti electron correlation formula, where the parameters were fit to the properties of the He atom. As for most parameterized electronic exchange-correlation functionals, the values of the parameters could significantly influence the final results. This parameter dependence will be investigated more thoroughly in future work.

This NEO-DFT/epc17 method is compared to the NEO-DFT/no-epc method, which includes no electron-proton correlation, as well as a grid-based method \cite{Marston_1989_3571} that is considered to be a reference for these electronically adiabatic systems. In the grid-based method, the total electronic energy is calculated using conventional electronic DFT/B3LYP for the hydrogen nucleus positioned at each grid point on a three-dimensional grid spanning the relevant region for the proton density, and the three-dimensional Schr{\"o}dinger equation is solved numerically for the proton using the Fourier grid Hamiltonian method \cite{Marston_1989_3571}. The proton density is defined to be the square of the proton wave function calculated in this manner. The grid method is numerically exact when the Born-Oppenheimer approximation is valid, as for these model systems.

\begin{figure}
	\begin{center}
		\includegraphics[scale=0.28]{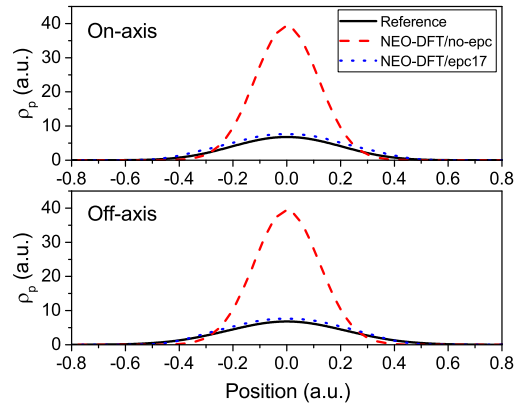}
		\caption{\label{fig:FHF}On-axis and off-axis proton densities for \ce{FHF-} calculated with the grid-based reference method(solid black line), the NEO-DFT/no-epc method (dashed red line), and the NEO-DFT/epc17 method (dotted blue line). The midpoint of the F---F bond is set as the origin.}
	\end{center}
\end{figure}

\begin{figure}
	\begin{center}
		\includegraphics[scale=0.28]{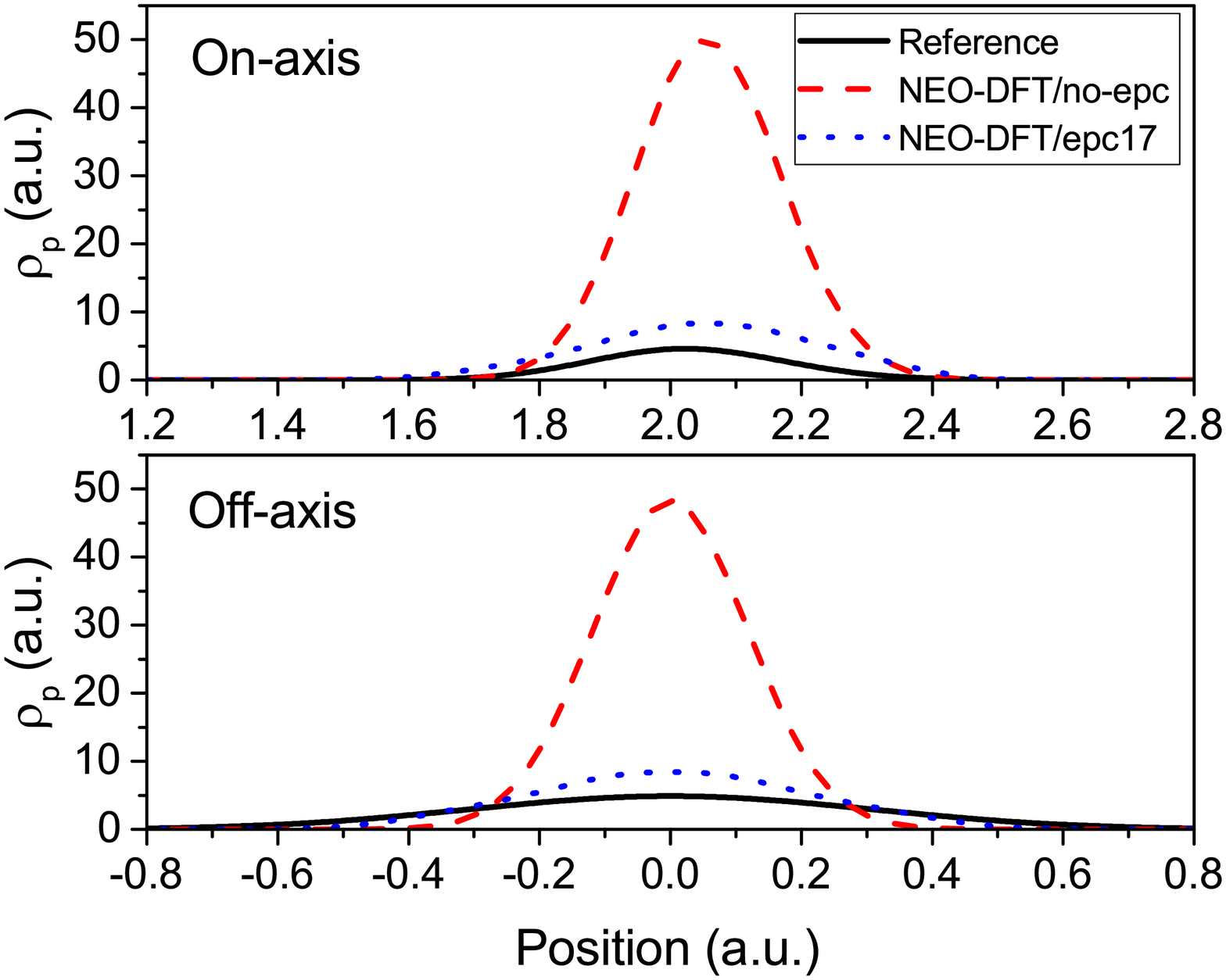}
		\caption{\label{fig:HCN}On-axis and off-axis proton densities for \ce{HCN} calculated with the grid-based reference method(solid black line), the NEO-DFT/no-epc method (dashed red line), and the NEO-DFT/epc17 method (dotted blue line). The position of the carbon atom is set as the origin. The expectation values for the proton coordinate on the principal axis are 2.001, 2.053, 2.028 a.u., respectively.}
	\end{center}
\end{figure}

Figure \ref{fig:FHF} shows the on-axis and off-axis proton densities for \ce{FHF-}, where the on-axis proton density is the slice along the axis connecting the two heavy nuclei, and the off-axis proton density is a slice perpendicular to this axis passing through the midpoint of the F---F bond. The NEO-DFT/no-epc method predicts a highly over-localized proton density, whereas the NEO-DFT/epc17 method is in excellent agreement with the grid-based reference method. Despite visual appearances, all of these proton densities are normalized in three-dimensional space, where multiplication by the volume element more heavily weights the regions further from the origin. In contrast, the results from previous papers were rescaled to be normalized in one dimension for the on-axis density \cite{Sirjoosingh_2015_214107, Culpitt_2016_44106, Brorsen_2015_214108} but contained significant errors because of a poor description of the off-axis proton density. The current epc17 functional corrects this problem and exhibits excellent agreement for both on-axis and off-axis proton densities with proper three-dimensional normalization. To our knowledge, no other electron-proton correlation functional can describe the proton densities with proper normalization in three dimensions even qualitatively accurately.

Without further parameterization, we applied the NEO-DFT/epc17 method to \ce{HCN}, an asymmetric case where the hydrogen atom is located at one end of the molecule. The results for this molecule are shown in Figure \ref{fig:HCN}. The NEO-DFT/epc17 method also provides greatly improved results compared to the NEO-DFT/no-epc method for both on-axis and off-axis proton densities, although minor differences can be observed. Note that altering the $a$, $b$, and $c$ parameters within the epc17 functional did not lead to significant improvement, implying transferability across different molecules for this functional, although additional molecules will be studied for further testing. Inclusion of the last two terms in Eq. \eqref{eq:Ecorr} may further improve this type of asymmetric case.

\begin{figure}
	\begin{center}
		\includegraphics[scale=0.45]{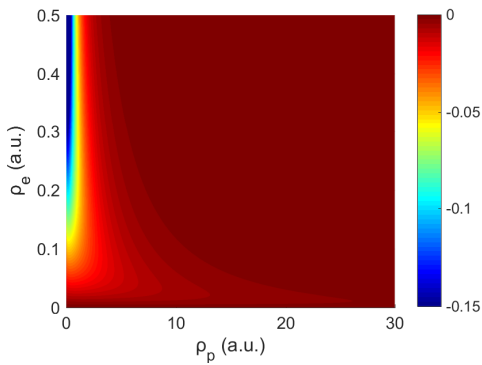}
		\caption{\label{fig:Vepc}Electron-proton correlation potential in the proton Kohn-Sham equation obtained from the epc17 functional as a function of the electron and proton densities. Because high proton density leads to high potential energy, the proton tends to become delocalized to reduce the potential energy.}
	\end{center}
\end{figure}

We investigated the reason for the more delocalized proton densities obtained with the epc17 functional by plotting the epc proton potential $v^{epc}_p$ as a function of the electron and proton densities in Figure \ref{fig:Vepc}. The ranges of the variables $\rho_e$ and $\rho_p$ are chosen to be in the physical regime for hydrogen atoms in a typical molecular environment. This plot shows that when the proton density increases, the epc potential for the proton also increases. Driven by this extra potential, the proton tends to become more delocalized to minimize the area with high proton density and hence decrease the potential energy. This property explains why such a simple LDA functional can effectively delocalize the proton density.

Thus, this LDA type of electron-proton correlation functional should be viewed as a first important step, with the understanding that gradient corrections will be critical for future functional development. Moreover, because the high and low density limits of electron-proton correlation are unknown, the parameterized form developed herein focuses on the physically meaningful regions for molecular systems, and the asymptotic high and low density limits are most likely not correct. Further analysis of these limits and development of functional forms that satisfy such constraints are other directions for future development.

In addition, similar to the LYP functional, the epc17 functional does not explicitly include the kinetic correlation effects.  These effects could be included with an adiabatic connection formula, \cite{Langreth_1975_1425, Imamura_2002_6458} as in a previous electron-proton correlation functional from our group. \cite{Sirjoosingh_2011_2689, Sirjoosingh_2012_174114}  However, because the epc17 functional is parameterized to fit numerically exact densities or energies, these effects are implicitly included in a semiempirical manner.  Another challenge in the development of electronic functionals has been the accurate inclusion of exchange terms for electrons. In contrast to electronic functionals, however, electron-proton functionals do not need to consider exchange between an electron and a proton because they are not identical particles.  Given the extensive literature aimed at improving electronic functionals, future efforts will explore alternative models and treatments for electron-proton correlation functionals.

\section{Conclusions}

In this paper, we developed an electron-proton correlation functional analogous to the LYP electron correlation functional based on the Colle-Salvetti formulation.  The implementation of this epc17 functional within the SCF procedure for multicomponent DFT produces accurate proton densities that are in good agreement with reference data for two representative molecules. To our knowledge, such accurate proton densities have not been achieved with any previous multicomponent DFT formulations. Moreover, the NEO-DFT/epc17 method is computationally inexpensive with the same formal scaling as conventional electronic DFT and is therefore promising for a wide range of future applications. For example, recent work based on the epc17 functional has been used to calculate proton affinities and to investigate the impact of proton delocalization on optimized geometries. \cite{Brorsen_2017_3488,Footnote} In addition, the NEO-DFT/epc17 method will enable the inclusion of nuclear quantum effects in calculations of p\textit{K}\textsubscript{a}'s, reaction paths, and reaction dynamics, as well as tunneling splittings and vibronic couplings. Therefore, this development lays the foundation for a wide range of potential new directions.

\begin{acknowledgments}
This work was supported by the National Science Foundation Grant No. CHE-1361293.	
\end{acknowledgments}

\section*{Supplementary material}
See supplementary material for the complete derivation of the epc17 functional.

\section*{References}

\end{document}